\documentclass[runningheads]{llncs}
\usepackage[T1]{fontenc}
\usepackage{graphicx}
\usepackage{color}
\usepackage{hyperref}
\usepackage{booktabs}
\usepackage{float}
\usepackage{tabularx}
\usepackage{needspace} 
\begin{document}
\title{\large From Course to Skill: Evaluating LLM Performance in Curricular Analytics}

\author{Zhen Xu\orcidID{0009-0004-3131-910X} \and
Xinjin Li\orcidID{0009-0007-1019-5023} \and
Yingqi Huan\orcidID{0009-0008-9304-9546} \and
Veronica Minaya\orcidID{0000-0002-0750-4805} \and
Renzhe Yu\orcidID{0000-0002-2375-3537}}
\authorrunning{Xu et al.}
\titlerunning{From Course to Skill: Evaluating LLM Performance in Curricular Analytics}

\institute{Columbia University, New York, NY, USA\\ 
\email{\{zx2393, xl3319, yh3755, vminaya, renzheyu\}@tc.columbia.edu}}
\maketitle              
\begin{abstract}
Curricular analytics (CA) -- systematic analysis of curricula data to inform program and course refinement -- becomes an increasingly valuable tool to help institutions align academic offerings with evolving societal and economic demands. Large language models (LLMs) are promising for handling large-scale, unstructured curriculum data, but it remains uncertain how reliably LLMs can perform CA tasks. In this paper, we systematically evaluate four text alignment strategies based on LLMs or traditional NLP methods for skill extraction, a core task in CA. Using a stratified sample of 400 curriculum documents of different types and a human-LLM collaborative evaluation framework, we find that retrieval-augmented generation (RAG) is the top-performing strategy across all types of curriculum documents, while zero-shot prompting performs worse than traditional NLP methods in most cases. Our findings highlight the promise of LLMs in analyzing brief and abstract curriculum documents, but also reveal that their performance can vary significantly depending on model selection and prompting strategies. This underscores the importance of carefully evaluating the performance of LLM-based strategies before large-scale deployment.
\end{abstract}

\section{Background}
\vspace{-2mm}
Curriculum is a core component of higher education, shaping students' intellectual growth and preparing students for the workforce, while also serving as a benchmark for program quality and institutional reputation. Given the rapid advancements in digital technology and the digital economy, institutions and educational stakeholders are increasingly seeking automated ways to analyze curriculum documents and generate evidence-based insights for improving curriculum design and delivery~\cite{pistilli2017guiding,chou2015open,hilliger2024curriculum}. In this context, curricular analytics (CA) has emerged as a subfield of learning analytics (LA), aimed at facilitating data-driven decision making and improvement in courses and programs~\cite{hilliger2024curriculum}. 

Despite its promise, CA remains relatively underdeveloped~\cite{yu2021}, due not only to a historical lack of digital curricula data but also to technical challenges of analyzing texts in a scalable and reliable manner. Curricula documents, such as course catalogs, syllabi, and reading materials, vary widely in structure, granularity, and language use, which makes automation difficult. Recent advances in natural language processing (NLP) have helped CA progress from rule-based to more sophisticated semantic approaches~\cite{kawintiranon2016,gottipati2018competency,javadian2024course,noveski2024,jovanovic2024curriculum,ehara2023,tan2023}, but major challenges still exist, including the difficulty in extracting fine-grained curricular constructs, the lack of standardized curricular ontologies, and the need for pedagogically grounded reasoning. With these challenges, achieving automated extraction of meaningful insights from curricula is still a considerable hurdle.

Recent advances in large language models (LLMs) offer new possibilities for curricular analytics. Their ability to efficiently extract the semantics of natural language could improve how we analyze and interpret curricular content and help identify complex educational constructs and curricular elements that were previously hard to capture. In addition, their natural language interfaces lower technical barriers, making CA more accessible to educators and researchers without extensive technical expertise. As a result, increasing efforts have been made to incorporate LLMs as analytical tools in CA research~\cite{zamecnik2024mapping,ehara2023,malik2024,kwak2024,li2024,noveski2024,nguyen2024rethinking}. While these studies have shown some promise of LLM-assisted CA, the reliability and generalizability of this promise across different curricular contexts are still not well understood. 

In this study, we systematically evaluate the performance of LLMs versus traditional NLP methods in the context of skill extraction, a core CA task that assesses how well course content aligns with workforce demands through the lens of skills. Skills are essential components of jobs and play a key role in shaping individuals' career outcomes in the labor market. Therefore, systematically examining skills and how they are developed through education is crucial for understanding students' future career trajectories and broader workforce trends~\cite{woessmann2024skills,deming2017growing}. By conducting this evaluation, our contributions are twofold. First, we provide one of the first systematic empirical assessments of LLMs' capabilities in curricular analytics, benchmarking their performance against major traditional NLP paradigms commonly used in CA. Second, we examine how LLM performance varies across different prompting strategies, model selections, and curriculum document types, offering practical guidelines and important considerations for researchers and practitioners seeking to integrate LLMs into CA research and applications.
\vspace{-2mm}

\section{Data and Methods}
\vspace{-2mm}

\subsection{Datasets}
\vspace{-1mm}
\subsubsection{Curriculum Documents.}
The curriculum documents used in this study come from two sources: (1) Course Syllabi from Open Syllabus\footnote{https://www.opensyllabus.org/}, a nonprofit archive of over 20.9 million higher education syllabi worldwide; (2) General catalog with short descriptions of individual courses from a large, urban, public two-year college in the United States, which is publicly available on its official website. We restrict our analysis to courses from the 2017–18 academic year for consistency and use stratified sampling to select 100 curriculum documents from each source, covering a diverse range of major areas and document length categories.

From stratified sampling, we generate four types of curriculum documents commonly used in CA: (1) course descriptions from the general catalog, (2) course descriptions in syllabi, (3) learning outcomes in syllabi, and (4) the combination of course descriptions and learning outcomes in syllabi. More details about the dataset can be found in the Supplemental Information.
\vspace{-4mm}

\subsubsection{Skill Framework.} \textit{O*NET} (Occupational Information Network) is a comprehensive database of job characteristics and worker skills from the US Department of Labor, which has been widely used for labor market analysis, curriculum design, and career guidance~\cite{hilton2010database,javadian2024course,burrus2013identifying,chauhan2019occupation}. We use the \textit{Detailed Work Activity (DWA)} taxonomy from \textit{O*NET}, which includes 2,070 short descriptions of real-world work activities across various occupations, and treat DWAs as skills in our analyses.

\vspace{-2mm}

\subsection{Skill Extraction}
A course can cultivate multiple skills, and here we extract the top 10 most relevant skills from each curriculum document for consistency of comparison. Skill relevance is measured by semantic alignment between a skill and a course. We apply the following four text alignment strategies.
\vspace{-4mm}
\subsubsection{Token-Based (TF-IDF):} Following~\cite{light2024student}, we calculate alignment scores between each course and DWA skill using TF-IDF weights, combined with relevance weights based on token importance in the DWA dataset relative to Wikipedia. The weighted TF-IDF scores are summed and normalized by token count to adjust for course length. 
\vspace{-4mm}
\subsubsection{Embedding-Based (BERT):} We adapt the embedding-based matching method from~\cite{javadian2024course}. SBERT, a siamese network-based model known for effective sentence embeddings~\cite{reimers2019sentence}, is used to calculate alignment scores between each curriculum document and skill description via cosine similarity.
\vspace{-4mm}
\subsubsection{Zero-Shot Prompting (ZERO-SHOT):} We use both open-source and proprietary models—GPT-4o, Llama 3.3-70B, Gemini 1.5 Pro, and Claude 3.5 Sonnet—to perform zero-shot skill extraction. Each model is prompted with the curriculum document and the predefined skill description list to perform skill extraction. The full prompt is provided in the Supplementary Information.
\vspace{-4mm}
\subsubsection{Retrieval-Augmented Generative (RAG):} We use RAG, a strategy that improves performance by retrieving relevant external information before generation~\cite{zamecnik2024mapping}, as a pre-filtering step to narrow the skill pool. Specifically, we embed 2,070 skill descriptions into a vector database and retrieve the top 20 most relevant skills\footnote{The number 20 aligns with the typical number of DWAs associated with each occupation in \textit{O*NET}.} based on cosine similarity with the curriculum document. These retrieved skills, along with the curriculum document and query, are then used to construct a structured prompt for the LLM to extract skills.
\vspace{-2mm}

\subsection{Performance Evaluation}
To evaluate the performance of each text alignment strategy, we score the alignment of extracted skills on a 0–5 scale using a human–LLM collaborative evaluation framework, and aggregate the scores to assess overall performance.

\begin{figure}
\centering
\vspace{-5mm}
\includegraphics[width=0.8\textwidth]{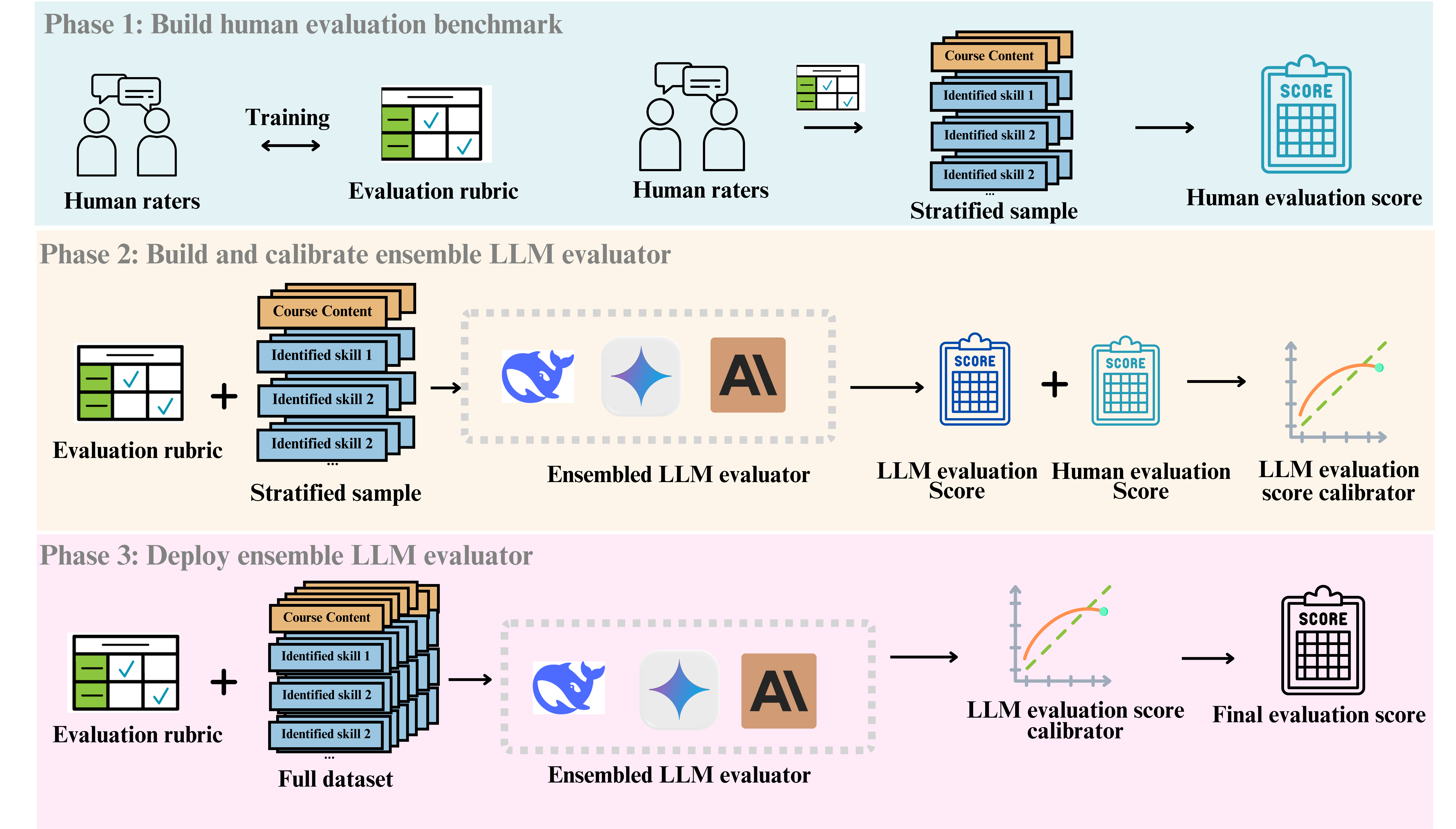}
\vspace{-3mm}
\caption{Human-LLM collaborative evaluation framework} \label{fig-evaluation}
\vspace{-5mm}
\end{figure}
\subsubsection{Human–LLM Collaborative Evaluation Framework.}
To assess the actual alignment of each extracted skill in a scalable manner, we build a human-LLM collaborative evaluation framework that combines human expertise and the power of LLMs, including three phases:
\vspace{-0.5mm}
\begin{enumerate}
    \item \textbf{Build human evaluation benchmark.} Three annotators first score skills extracted from 10 courses across all methods to iteratively refine and finalize a scoring rubric (Table~\ref{rubric}). Two annotators then apply the rubric to skills from 60 randomly selected curricula, resolving discrepancies through discussion and re-scoring until acceptable reliability was achieved (Cohen’s Kappa = 0.747, ICC = 0.862). Finally, the two annotators score skills from another 60 stratified samples across data types, subjects, and alignment strategies, again resolving any differences to produce the finalized human evaluation benchmark dataset.
    \item \textbf{Build and calibrate ensemble LLM evaluator.} Drawing on LLM-as-a-judge frameworks in the NLP domain~\cite{wang2022self}, we use several state-of-the-art reasoning models—DeepSeek R1, Gemini 1.5 Pro, and Claude 3.5 Sonnet—as an ensemble evaluator. Each model scores the skills using few-shot prompts based on our rubric, and their scores are averaged. To improve reliability, we train a calibration model that predicts human scores from LLM outputs using linear regression with quantile interpolation. This helps correct bias and aligns the distribution of LLM predictions with human evaluations. The human evaluation benchmark dataset is split 80\% and 20\% for training and testing, using 10-fold cross-validation.  The calibrated model achieves an accuracy of 0.709, a weighted Cohen’s Kappa of 0.767, and a Krippendorff’s alpha of 0.761, indicating good consistency with human judgments.
    \item \textbf{Deploy ensemble evaluator.} Lastly, we deploy the calibrated LLM evaluator to assess the top 10 skills identified by each of the 10 alignment methods across 400 curriculum samples. Each extracted skill is independently scored using few-shot prompting, and the ensemble outputs are then calibrated using the model trained in Phase 2.
\end{enumerate}

\begin{table}[h!]
\vspace{-5mm}
\raggedleft
\footnotesize
\setlength{\tabcolsep}{6pt} % tighter columns
\renewcommand{\arraystretch}{0.95} % tighter rows
\begin{tabularx}{\textwidth}{>{\centering\arraybackslash}p{0.04\textwidth} X}
\toprule
\centering Score & Criteria \\
\midrule
\centering 5 & Core learning objective of the course; explicitly covered.\\
\centering 4 & Aligns with the course; students should be able to perform it after completion.\\
\centering 3 & Not explicitly covered, but transferable skills may be developed.\\
\centering 2 & Within the same domain, but not directly relevant.\\
\centering 1 & Outside the scope of the course; belongs to a different domain.\\
\bottomrule
\end{tabularx}
\caption{Rubric for evaluating the actual alignment of each extracted skill}\label{rubric}
\vspace{-5mm}
\end{table}

\vspace{-8mm}
\subsubsection{Performance Metrics.}
The skill alignment scores generated above are further aggregated into four metrics to evaluate the overall performance of each text alignment strategy: (1) $Precision_{5}$: \% of top 10 extracted skills that score 5; (2) $Precision_{4}$: \% top 10 extracted skills that score 4 or higher; (3) $Mean$: Average alignment score of the top 10 skills; (4) Normalized Discounted Cumulative Gain$(NDCG)$: A metric that evaluates the ranking accuracy of information retrieval systems, with scores ranging from 0 to 1. Higher values indicate more accurate rankings~\cite{busa2012apple}.
\vspace{-2mm}
\section{Results}
\vspace{-2mm}
Table~\ref{table-Overall performance} summarizes the overall performance of each alignment method across the full dataset. RAG consistently outperforms both zero-shot and traditional methods in terms of extraction precision. Among zero-shot prompting methods, only the best-performing model, GPT-4o, surpasses traditional NLP approaches, while the average performance of zero-shot methods remains lower than traditional methods.
\vspace{-5mm}
\begin{table}
\centering
\footnotesize
%\linespread{1}
\begin{tabular}{lllllllllll}
\toprule
 & TF-IDF & BERT & \multicolumn{4}{c}{ZERO-SHOT} & \multicolumn{4}{c}{RAG} \\
\cmidrule(lr){2-2}\cmidrule(lr){3-3}\cmidrule(lr){4-7} \cmidrule(lr){8-11}
 &  &  & GPT & Llama  & Claude & Gemini & GPT & Llama  & Claude & Gemini \\
\midrule
$Precision_{5}$ & 0.100  & 0.043  & 0.244  & 0.032 & 0.032 & 0.055 & 0.540 & 0.432 & 0.416 & 0.432 \\
$Precision_{4}$ & 0.269 & 0.240 & 0.418 & 0.116 & 0.160 & 0.199 & 0.820 & 0.715 & 0.695 & 0.721 \\
$Mean$ & 2.418 & 2.344 & 2.981 & 1.824 & 2.074 & 2.216 & 4.268 & 3.985 & 3.946 & 3.993 \\
$NDCG$ & 0.887 & 0.878 & 0.868 & 0.881 & 0.869 & 0.899 & 0.959 & 0.971 & 0.973 & 0.973 \\
\bottomrule
\end{tabular}
\caption{Overall performance across the entire dataset}
\label{table-Overall performance}
\vspace{-5mm}
\end{table}

\begin{figure}
\vspace{-10mm}
\includegraphics[width=\textwidth]{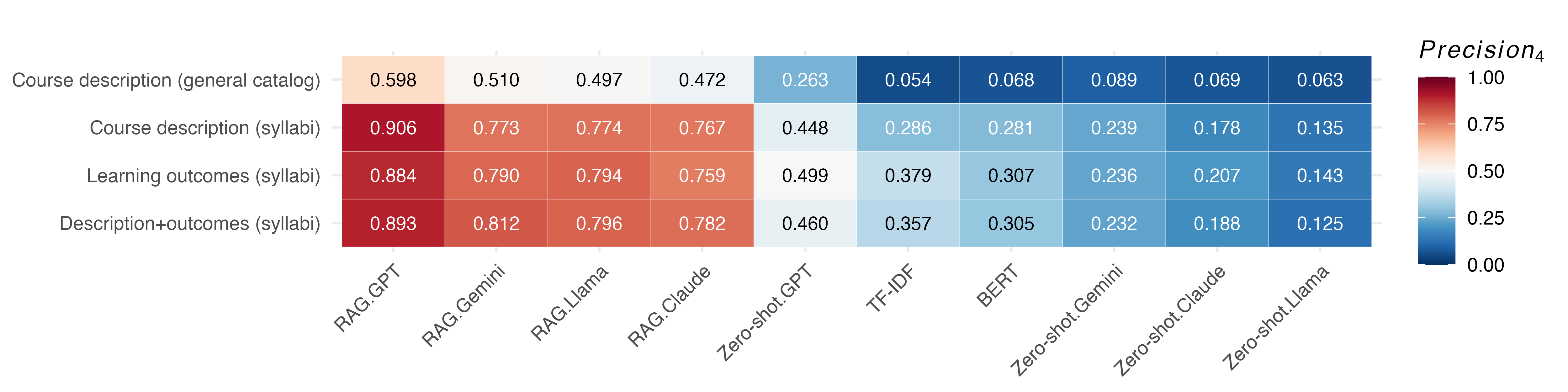}
\vspace{-10mm}
\caption{$Precision_4$ comparison across different types of curriculum documents} \label{fig-rq2_1}
\vspace{-9mm}
\end{figure}

We further examine performance heterogeneity across different types of curriculum documents, focusing on $Precision_{4}$, as identifying relevant skills is typically prioritized in skill extraction research. As shown in Figure \ref{fig-rq2_1}, RAG consistently outperforms both traditional NLP and zero-shot methods across all curriculum document types. In general catalogs, RAG achieves 47.2–59.8\% precision, correctly identifying about half of the top 10 skills. In contrast, traditional NLP methods and most zero-shot models (except GPT-4o) score below 10\%, often missing all relevant skills. For syllabi, traditional methods like TF-IDF and BERT perform better than most zero-shot models, identifying around 2.8–3.8 relevant skills, while others find only about 2.4. RAG again performs best, with 7.6–9.1 relevant skills on average.

\vspace{-2mm}
\section{Discussion and Conclusion}
\vspace{-2mm}
In this study, we present one of the first systematic evaluations of text alignment strategies for the skill extraction task in CA, comparing traditional NLP methods and LLM-based approaches across different types of curriculum documents. Our findings reveal key insights into the overall performance and generalizability of these methods.

LLM-based methods, especially those using RAG, consistently outperform traditional approaches in alignment quality, precision, and ranking accuracy. This advantage holds across all types of curriculum documents. In particular, LLMs show strong improvements in handling brief, general documents, such as open catalogs, where traditional methods often struggle due to limited detail and the pedagogical reasoning required to address granularity mismatches. These findings underscore the promise of LLMs in addressing longstanding challenges in CA, especially when dealing with sparse or heterogeneous educational data.

We also examined how LLM performance varies across different models and document types. In zero-shot settings, performance differed notably between open-source and proprietary models, as well as by model size, parameters, and optimization goals. However, using RAG helped reduce this variation, resulting in more stable outcomes across models. Additionally, zero-shot prompting worked better on the most difficult curriculum document types for traditional methods, like general catalogs, but was less effective on structured, information-rich documents.

Our findings have several practical implications. First, LLMs can be a powerful alternative to traditional NLP methods when working with low-information, highly summarized curriculum documents. Second, while LLMs are promising for empowering CA tasks, effective use requires careful design and tuning beyond zero-shot prompting alone. Third, given the performance variation across prompting and model selection, careful evaluation, transparent reporting of methodological choices, and validation of LLM-involved analyses are crucial to ensuring the rigor and trustworthiness of research conclusions.
\vspace{-2mm}

\section*{Acknowledgements}
\vspace{-2mm}
This work is supported by a Faculty Collaboration Grant from Teachers College, Columbia University. We also thank Joe Karaganis for providing access to Open Syllabus data.
\vspace{-2mm}

\section*{Supplemental Information}
\vspace{-2mm}
Supplemental Information can be accessed at: \url{https://github.com/AEQUITAS-Lab/Evaluation-of-LLM-in-CA-AIED-2025}
\vspace{-2mm}

\bibliographystyle{splncs04}
\bibliography{mybibliography}
\end{document}